\documentclass[sigconf]{acmart}

\usepackage{algpseudocode}
\usepackage{amsmath}
\usepackage{amsthm}
\usepackage{balance}  % for  \balance command ON LAST PAGE  (only there!)
\usepackage{bm}
\usepackage{breqn}
\usepackage{cleveref}
\usepackage{graphicx}
\usepackage{hyperref}
\usepackage{pgfplots}
\usepackage{relsize}
\usepackage{subcaption}
\usepackage{tablefootnote}
\usepackage{tikz}
\usepackage{xspace}
\usepackage{multirow}
\usepackage{caption}
\usepackage{pifont}
\usepackage{enumitem}
\usepackage{algorithm}
\usepackage{wrapfig}
\expandafter\def\csname ver@subfig.sty\endcsname{}
\usepackage{subfig}
\usepackage{xcolor}

\usetikzlibrary{arrows}
\usetikzlibrary{patterns}
\usetikzlibrary{positioning} 
\usetikzlibrary{fit}
\usetikzlibrary{shapes.geometric}
\usetikzlibrary{arrows.meta}
 \usetikzlibrary{calc} 
\tikzset{>=latex}
\usepgfplotslibrary{groupplots}

\definecolor{darkgreen}{RGB}{0,100,0}
\definecolor{lightgreen}{RGB}{204, 253, 197}
\definecolor{lightred}{RGB}{246, 193, 192}
\definecolor{lightblue}{RGB}{191, 191, 251}
\definecolor{lightyellow}{RGB}{255, 255, 198}
\definecolor{lightorange}{RGB}{247, 215, 157}

\definecolor{contrastpink}{RGB}{212, 17, 89}
\definecolor{contrastblue}{RGB}{26, 133, 255}

\makeatletter
\renewcommand\tagform@[1]{\maketag@@@{\ignorespaces#1\unskip\@@italiccorr}}
\makeatother

\newcommand{\sys}{ScanTwin\xspace}

\newcommand{\squishlist}{
   \begin{list}{$\bullet$}
    { \setlength{\itemsep}{0pt}
      \setlength{\parsep}{1pt}
      \setlength{\topsep}{6pt}
      \setlength{\partopsep}{0pt}
      \leftmargin=25pt
\rightmargin=0pt
\labelsep=5pt
\labelwidth=10pt
\itemindent=0pt
\listparindent=0pt
\itemsep=\parsep
    }
}
\newcommand{\squishend}{\end{list}}

\newcommand{\cut}[1]{}

\pgfplotsset{compat=1.18}

\definecolor{color1}{RGB}{216, 088, 032} %reddish orange
\definecolor{color2}{RGB}{127, 206, 205} %light blue (bright)
\definecolor{color3}{RGB}{78, 42, 132} %purple
\definecolor{color4}{RGB}{217, 200, 038} %gold
\definecolor{color5}{RGB}{ 88,185,71}  %bright green
\definecolor{color6}{RGB}{80, 145, 205} % dark blue

\copyrightyear{2026}
\acmYear{2026}
\setcopyright{cc}
\setcctype{by}
\acmConference[SeQureDB '26]{Workshop on Secure and Private Data Management}{May 31-June 05, 2026}{Bengaluru, India}
\acmBooktitle{Workshop on Secure and Private Data Management (SeQureDB '26), May 31-June 05, 2026, Bengaluru, India}
\acmDOI{10.1145/3807894.3810278}
\acmISBN{979-8-4007-2219-6/2026/05}

\begin{document}
\title[ScanTwin: Simulating Performance Regressions Without Access to Tenant Data]{ScanTwin: Simulating Performance Regressions \\
Without Access to Tenant Data}

%%
%% The "author" command and its associated commands are used to define the authors and their affiliations.
\author{Donghyun Sohn}
\affiliation{%
  \institution{Northwestern University}
  \city{Evanston}
  \state{Illinois}
  \country{USA}
}
\email{donghyun.sohn@u.northwestern.edu}

\author{Jennie Rogers}
\affiliation{%
  \institution{Northwestern University}
  \city{Evanston}
  \state{Illinois}
  \country{USA}
}
\email{jennie@northwestern.edu}

\begin{abstract}

In cloud data platforms, developers often encounter performance regressions that occur in specific tenant datasets. However, due to confidentiality constraints, they cannot access the original data, which makes it difficult to reproduce these regressions locally. Current methods for synthetic data usually focus on statistical properties, such as matching data distributions or improving query accuracy. However, they overlook the physical properties that control how the engine behaves during scans, including row-group pruning.

We propose \sys,  a lightweight framework that extracts a per-row-group sketch from the Parquet footer, including boundary values and compressed sizes, and releases them under $\varepsilon$-differential privacy using a boundary parameterization. On TPC-H and SSB (6M rows), \sys achieves 0\% pruning error and less than  1\% byte error at $\varepsilon{=}\infty$. Under $\varepsilon{=}5$, high-selectivity queries ($>$30\%) incur below 8.5\% pruning error on both datasets, and per-query scan timing on DuckDB closely tracks the original.
\end{abstract}
\maketitle

\section{Introduction}
\label{sec:intro}

Cloud data platforms often experience performance regressions that affect only specific tenant datasets. Diagnosing such regressions requires reproducing the original execution behavior, but privacy rules prevent the development team from accessing the original data. Therefore, they need synthetic data that triggers the same engine behavior as the original.

The main reason current synthetic data fails to reproduce performance is that it ignores the physical layout of files. Most existing methods~\cite{ge2024privacyenhanced, mckenna2022aim, zhang2017privbayes} focus only on statistical fidelity, like matching data distributions or query counts. But in columnar engines like DuckDB, the scan operator reads per-row-group (RG) statistics from the Parquet footer and skips any RG whose value range does not match the query predicate. If the synthetic file does not have the same physical arrangement of values, the engine will scan every RG even if the authentic data would have pruned them. The key problem is not just the values themselves, but where they are partitioned across RGs.

In this paper, we focus on scan-level behavior because the scan operator determines which RGs are read from storage, bounding the I/O cost of all downstream operators. Since joins and aggregations require additional state that cannot be recovered from a Parquet footer, we defer them to future work.

To solve this, we develop ScanTwin, a lightweight framework that uses the Parquet footer. ScanTwin extracts a per-RG sketch of the file, including min/max boundaries and compressed sizes, and we protect this information with $\varepsilon$-differential privacy (DP)~\cite{dwork2006calibrating}. Boundary parameterization is a key idea that reduces $2K$ min/max values to $K-1$ interior values, maintaining the accuracy of the synthetic file while minimizing DP noise. Our contributions are:
\begin{itemize}[leftmargin=*,nosep]
  \item We formalize engine reaction equivalence using RGs scanned and bytes read as proxies for scan-level performance, enabling developers to reproduce and diagnose regressions without accessing the original data~(\S\ref{sec:problem}).
  \item We design a boundary parameterization for the DP sketch that fixes the outer limits and only noises the $K{-}1$ interior points, with $\varepsilon$-DP guaranteed via the bounded Laplace mechanism (\S\ref{sec:dp}).
  \item On TPC-H and SSB, \sys closely reproduces per-query scan timing on DuckDB. At $\varepsilon{=}5$, pruning error for high-selectivity queries stays below 8.5\% for both datasets~(\S\ref{sec:experiment}).
\end{itemize}
\section{Problem}
\label{sec:problem}

\noindent\textbf{Setting.}
Given a confidential Parquet dataset $D$, a scan-heavy 
workload $W$, and a privacy budget $\varepsilon$, we extract a compact execution-relevant sketch $\mathcal{S} = \textsc{Sketch}(D)$ inside the trusted environment. The sketch is then released with $\varepsilon$-DP noise and used to generate $D'$ on an untrusted machine.

\smallskip\noindent\textbf{Goal (Engine Reaction Equivalence).}
For each query $q \in W$, the physical execution profile on $D'$ should approximate that on $D$: $\textsc{Profile}(q, D) \approx \textsc{Profile}(q, D')$.
Here, $\textsc{Profile}$ captures scan-level signals such as the number of RGs scanned and the bytes read. By reproducing these signals, \sys ensures that the engine makes the same pruning decisions and incurs similar I/O volume as on the original data. In practice, high-selectivity queries dominate the workload's cost: a query scanning 40-60 RGs incurs substantially more I/O than one scanning 1-2 RGs. \sys therefore focuses on these high-cost queries, where regressions have a large impact.

\section{Synthesis}
\label{sec:synthesis}

The \sys framework follows a three-stage process: (1)~sketch extraction from the Parquet footer, (2)~DP noise application, and (3)~synthetic Parquet generation.

% --- 3.1 ---
\subsection{Sketch Extraction}
\label{sec:sketch}

Given a Parquet file sorted on the filter column $c$, we read only the file footer, not the data pages, and record three values per RG~$i$:
 
\begin{itemize}[leftmargin=*,nosep]
   \item $\mathsf{min}_i, \mathsf{max}_i$:
        the minimum and maximum of column $c$ in RG~$i$, read from the Parquet footer statistics.
  \item $\mathsf{size}_i$:
        the total compressed size (bytes) of RG~$i$.
\end{itemize}

\begin{figure}[t]
  \centering
  \includegraphics[width=\columnwidth]{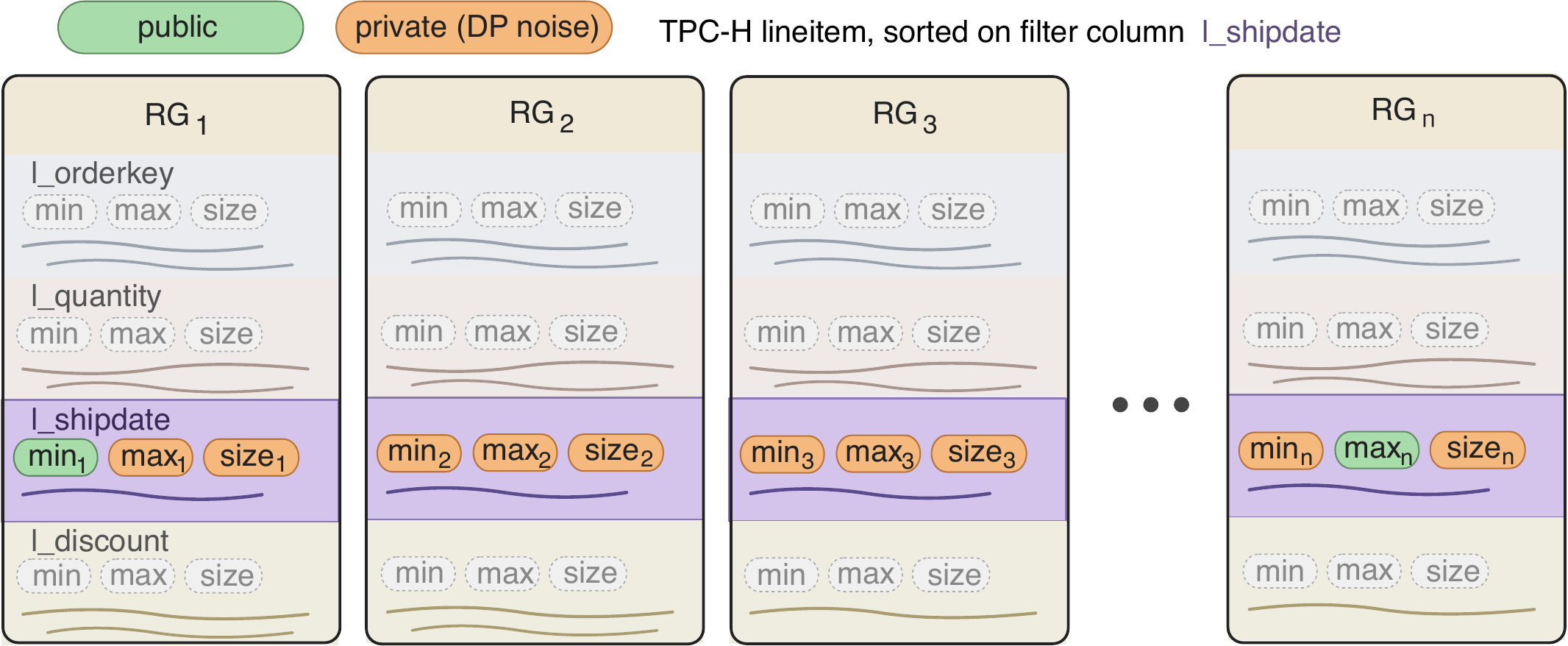}
  \vspace{-1em}
  \caption{\sys's sketch over a Parquet file.}
  \label{fig:layout}
\end{figure}

Figure~\ref{fig:layout} illustrates this layout. Only the filter column's per-RG metadata is extracted; other columns are ignored at sketch time. The sketch $\mathcal{S}$ is composed of the row count $N$, the number of RGs $K$, and the column count. Since Parquet stores statistics in the footer, we extract the sketch without reading data pages. Sorting by the filter column is standard practice (e.g., Snowflake clustering keys~\cite{snowflake_clustering}, Databricks Z-order~\cite{databricks_zorder}), and is necessary for effective zone-map pruning, the target behavior \sys reproduces.

We treat the compressed storage size $\mathsf{size}_i$ as the sensitive information that requires protection through DP. For the Parquet writer, we use a fixed row count for each RG ($RG\_size = n$), following DuckDB’s recommendation~\cite{duckdb_parquet} of 100K--1M rows. While $n$ is public (derived from $N$ and $K$), the compressed size
$\mathsf{size}_i$ is sensitive as it depends on the entropy of the tenant data.

\subsection{Differentially Private Sketch}
\label{sec:dp}

\smallskip
\noindent\textbf{Public metadata and trust model.}
Sketch extraction and noise injection occur within a trusted boundary managed by the tenant. Once the noise is added, the sketch can be safely shared with anyone. This is because it contains only per-RG aggregate statistics, not individual records, and the synthetic file is a deterministic function of the noisy sketch, revealing no additional information beyond what the DP release already protects.

We assume the following as public:
(1)~The domain of the filter column $[d_L, d_U]$.
(2)~The query workload $W$.
(3)~The maximum multiplicity $m$ (the highest number of times any single value appears). If $m$ is unknown, we can safely use the row count per RG ($m = n$). Only the actual data $D$ is confidential. Our method provides pure $\varepsilon$-DP, secure against a computationally unbounded adversary.

\begin{table}[t]
\centering
\small
\caption{Sketch contents.}
\vspace{-0.5em}
\label{tab:sketch}
\begin{tabular}{@{}lll@{}}
\toprule
\textbf{Level} & \textbf{Feature} & \textbf{Role} \\
\midrule
File-level
  & $N$ (total rows)              & Scan volume \\
  & $K$ (number of RGs), RG size  & Pruning granularity \\
  & Column count, codec           & I/O width, decode cost \\
\midrule
Per-RG
  & $\mathsf{min}_i, \mathsf{max}_i$ (filter col) & Pruning decisions \\
  & $\mathsf{size}_i$ (compressed bytes) & Per-RG I/O cost \\
\bottomrule
\end{tabular}
\end{table}
\vspace{-0.5em}

\smallskip
\noindent\textbf{Neighboring relation.}
Neighboring datasets are defined as those that differ by the value of a single row within a fixed positional RG assignment. Rows are assigned to RGs by position (rows $1 \ldots n$ to RG~1, rows $n{+}1 \ldots 2n$ to RG~2, and so forth), and a neighboring dataset replaces the value of one row while preserving its RG assignment. Because RGs are disjoint, modifying a single row affects only the min, max, and size of that specific RG.

\smallskip
\noindent\textbf{Boundary parameterization.}
In a sorted file, boundaries of neighboring RGs are usually similar, with $\mathsf{max}_{i-1} \approx \mathsf{min}_{i}$. Instead of noising $\mathsf{min}$ and $\mathsf{max}$ separately for every RG and then repairing inconsistencies, we use a single boundary-based representation. We fix the outermost points to the public domain limits, namely $\beta_0 = d_L$ and $\beta_K = d_U$, and treat only the $K-1$ internal boundaries as parameters, with each $\beta_i$ corresponding to $\mathsf{max}_i$. In this notation, $i$th RG is $\mathrm{RG}_i=[\beta_{i-1},\,\beta_i]$. Since the endpoints are already public, they do not need to be perturbed, so noise is applied only to the internal boundaries. This lowers the number of noisy values from $2K$ to $K-1$. It also enables an even split of the privacy budget across all RGs.

\smallskip
\noindent\textbf{Privacy budget.}
Since the $K$ RGs are disjoint subsets, parallel composition~\cite{dwork2014dp} lets every RG use the full privacy budget $\varepsilon$. Within each RG, we release two statistics ($\beta_i$ and $\mathsf{size}_i$) and split the budget evenly via sequential composition~\cite{dwork2006calibrating}, allocating $\varepsilon/2$ to each.

\smallskip
\noindent\textbf{Sensitivity.}
The sensitivity of an interior boundary $\beta_i$ represents the maximum possible shift caused by changing a single row. If the data has duplicates, changing one row can move the boundary by at most the number of times that specific value appears ($m$). Since boundaries are constrained to the public domain, we cap sensitivity at the domain width: $\Delta_\beta = \min(m,\, d_U - d_L)$. In practice, $m$ is conservative for skewed distributions; tighter bounds via DP histograms~\cite{xu2013differentially} are deferred to future work. For the compressed size, changing one row affects the RG’s size by at most the size of one uncompressed row:
$\Delta_s = \frac{\sum_{c \in \mathrm{Cols}} \mathsf{UncompressedSize}(c)}{n}$.

\begin{algorithm}[t]
\small
\caption{DP Sketch Release}
\label{alg:dp}
\begin{algorithmic}[1]
\Require Per-RG metadata $\{(\mathsf{min}_i, \mathsf{max}_i,
         \mathsf{size}_i)\}_{i=1}^{K}$, budget $\varepsilon$,
         public domain $[d_L, d_U]$, max multiplicity $m$
\State $\varepsilon_{\beta} \gets \varepsilon/2$; \quad $\varepsilon_{s} \gets \varepsilon/2$ \Comment{Budget splitting}
\State $\Delta_\beta \gets \min(m,\, d_U - d_L)$; \quad $\Delta_s \gets \mathsf{UncompressedRowSize}$
\State $b^* \gets \mathsf{FixedPoint}(\Delta_\beta,\, d_L,\, d_U,\, \varepsilon_{\beta})$ \Comment{Via Holohan et al.~\cite{holohan2018boundedlaplace}}
\State $\beta_0 \gets d_L$; \quad $\beta_K \gets d_U$

\For{$i = 1 \ldots K-1$}
    \State $\tilde{\beta}_i \gets \mathsf{BoundedLaplace}(\beta_i,\, b^*,\, d_L,\, d_U)$ \Comment{Boundary release}
\EndFor

\For{$i = 1 \ldots K$}
    \State $\tilde{\mathsf{size}}_i \gets \mathsf{BoundedLaplace}(\mathsf{size}_i,\, \Delta_s/\varepsilon_{s},\, 0,\, \infty)$ \Comment{Size release}
\EndFor

\State \textbf{sort} $(\tilde{\beta}_1, \ldots, \tilde{\beta}_{K-1})$ \textbf{and return} $\{(\tilde{\beta}_{i-1},\,\tilde{\beta}_i,\,\tilde{\mathsf{size}}_i)\}_{i=1}^{K}$ \Comment{Post-processing}
\end{algorithmic}
\end{algorithm}

\smallskip
\noindent\textbf{Noise mechanism (Algorithm~\ref{alg:dp}).}
We release each interior boundary using the bounded Laplace mechanism of Holohan et al.~\cite{holohan2018boundedlaplace}, which samples from a Laplace distribution conditioned on the public domain $[d_L, d_U]$ via inverse CDF transform. The mechanism's noise scale $b^*$ is computed via the fixed-point iteration of~\cite[Theorem 4.4]{holohan2018boundedlaplace}, which yields the smallest $b$ guaranteeing pure $\varepsilon$-DP. When $\Delta_\beta = d_U - d_L$, this reduces to the standard Laplace scale $b^* = b_0 = \Delta_\beta / \varepsilon_\beta$ exactly~\cite[Lemma 4.2] {holohan2018boundedlaplace}. Otherwise, $b^*$ is modestly larger. In our experiments, the noise inflation $b^*/b_0$ is exactly $1.0$ on TPC-H (where the cap binds at $\Delta_\beta = d_U - d_L$) and remains below $1.009$ on SSB across all  $\varepsilon \geq 0.01$. Compressed sizes are released analogously on $[0, \infty)$, where the inflation is negligible since size values are far from the lower bound.

After noising, we sort $(\tilde{\beta}_1, \ldots, \tilde{\beta}_{K-1})$ in ascending order to restore monotonicity. Sorting is a deterministic function of the noisy boundaries and incurs no extra privacy cost. The outer boundaries $\beta_0 = d_L$ and $\beta_K = d_U$ are public constants and require no repair.

\subsection{Synthetic Parquet Generation}
\label{sec:synth}

Using the noisy sketch, we build a replacement Parquet file. This new file is designed to trigger the same pruning results and I/O workload as the original data.

\smallskip
\noindent\textbf{Step 1: Filter column.}
For each RG~$i$, we pick $n$ values uniformly from the estimated domain $[\tilde{\beta}_{i-1},\, \tilde{\beta}_i]$ and sort within the group. This
ensures the Parquet writer populates the footer with min/max statistics, so the engine's zone-map pruning yields skip-or-scan decisions that closely match those in the original.

\smallskip
\noindent\textbf{Step 2: Padding columns.}
We add $C{-}1$ padding columns of random bytes to match each RG's target compressed size. Because ZSTD overhead varies with payload, we use a two-pass calibration: write an initial estimate (with compressed size $\mathit{base}_i$ from the filter column alone), measure the output, then adjust the padding to reach the noisy target $\tilde{\mathsf{size}}_i$. This uses only the noisy sketch and codec behavior, consuming no additional privacy budget: $\mathit{pad}_i = \max\bigl(0,\;\tilde{\mathsf{size}}_i - \mathit{base}_i\bigr)$.
 
\smallskip
\noindent\textbf{Step 3: Write.}
The synthetic file uses the same RG size and ZSTD codec as the original. The synthetic data is semantically neutral. Dummy values are random, and filter column values only need to fall within the estimated range. Our main goal is to ensure the query engine maintains identical pruning decisions and I/O volumes. 

% --- Table 2 ---
\begin{table}[t]
\centering
\small
\caption{Full comparison at $\varepsilon{=}\infty$ (no privacy).}
\label{tab:comparison}
\begin{tabular}{@{}llrr@{}}
\toprule
\textbf{Dataset} & \textbf{Method} & \textbf{MAPE-RG} & \textbf{MAPE-Bytes} \\
\midrule
TPC-H   & Random          & 3{,}631\% & 5{,}274\% \\
        & Marginal        & 3{,}616\% & 5{,}274\% \\
        & Sorted-Global   & 28.5\%    & 88.4\%    \\
        & \sys-MinMax     & 0.0\%     & 7.9\%     \\
        & \sys-Full       & 0.0\%     & 0.2\%     \\
\midrule
SSB     & Random          & 1{,}421\% & 1{,}760\% \\
        & Marginal        & 1{,}431\% & 1{,}818\% \\
        & Sorted-Global   & 0.0\%     & 35.1\%    \\
        & \sys-MinMax     & 0.0\%     & 10.4\%    \\
        & \sys-Full       & 0.0\%     & 0.2\%     \\
\bottomrule
\end{tabular}
\end{table}

\section{Experimental Results}
\label{sec:experiment}

We evaluate \sys on:
(1)~pruning and I/O fidelity compared to layout-unaware baselines,
(2)~the privacy--utility tradeoff,
(3)~engine-level validation, and
(4)~sensitivity to the multiplicity bound~$m$.

\subsection{Setup}

We assess \sys using two standard analytical benchmarks:
TPC-H \texttt{lineitem} (6M rows, filter on \texttt{l\_shipdate})~\cite{tpc-h} and SSB \texttt{lineorder} (6M rows, filter on \texttt{lo\_orderdate})~\cite{ssb}. Both are sorted by the filter column and written as Parquet with 100K rows per RG (61 RGs each) and ZSTD compression. All experiments use \texttt{WHERE col$\leq$cutoff} queries at selectivities log-spaced from 0.1\% to 95\%. For privacy analysis, we report mean $\pm$ std over five seeds (42, 7, 2025, 777, 867). Engine validation runs on DuckDB (v1.4.4), single-thread, 4\,GB RAM, and reports median of five runs per query.

\smallskip
\noindent\textbf{Metrics.}
\textbf{MAPE-RG}: mean absolute percentage error in RGs scanned (pruning fidelity).
\textbf{MAPE-Bytes}: error in compressed bytes read (I/O fidelity).
For engine validation, we report per-query scan times and total workload time ratio (synthetic/original). 
% \donghyun{We treat scan timing as the ground-truth utility metric, since it directly reflects regression-relevant performance, and MAPE-RG as a sensitive diagnostic that surfaces small layout differences before they manifest in timing.}

\smallskip
\noindent\textbf{Methods.}
We compare against Random (unsorted uniform), Marginal (global histogram), Sorted-Global (globally sorted, no per-RG metadata), \sys-MinMax (per-RG bounds, uniform padding), and \sys-Full (per-RG bounds, two-pass calibrated padding). All methods share the same schema, row count, codec, and RG size.

\subsection{Layout-Aware vs.\ Layout-Agnostic Synthesis}
\label{sec:without-dp}

Table~\ref{tab:comparison} isolates layout reproduction from DP noise ($\varepsilon{=}\infty$).
Random and Marginal scan all 61 RGs (MAPE-RG $>$1{,}400\%). Both \sys variants achieve 0\% MAPE-RG. Sorted-Global reaches 28.5\% on TPC-H; its 0\% on SSB is coincidental, as the near-uniform date distribution matches uniform sampling. \sys-Full's 0.2\% MAPE-Bytes vs.\ Sorted-Global's 35.1\% on SSB confirms that per-RG size calibration is essential for I/O fidelity.

\subsection{Privacy-Utility Tradeoff}
\label{sec:with-dp}
\begin{figure}[t]
  \centering
  \includegraphics[width=0.8\columnwidth]{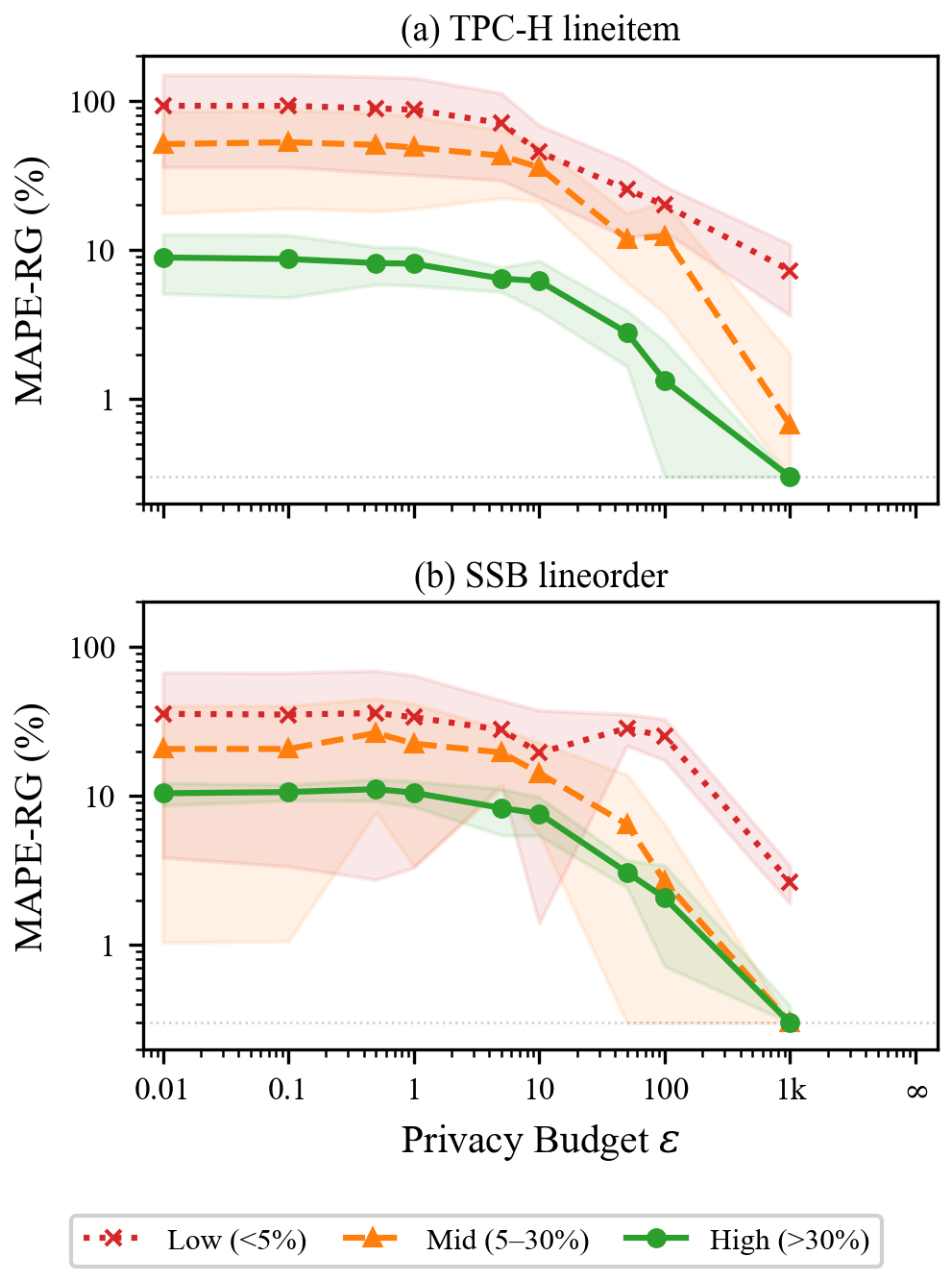}
  \vspace{-1em}
  \caption{MAPE-RG vs.\ $\varepsilon$ broken down by selectivity.}
  \label{fig:epsilon}
\end{figure}

\begin{figure}[t]
  \centering
  \includegraphics[width=0.8\columnwidth]{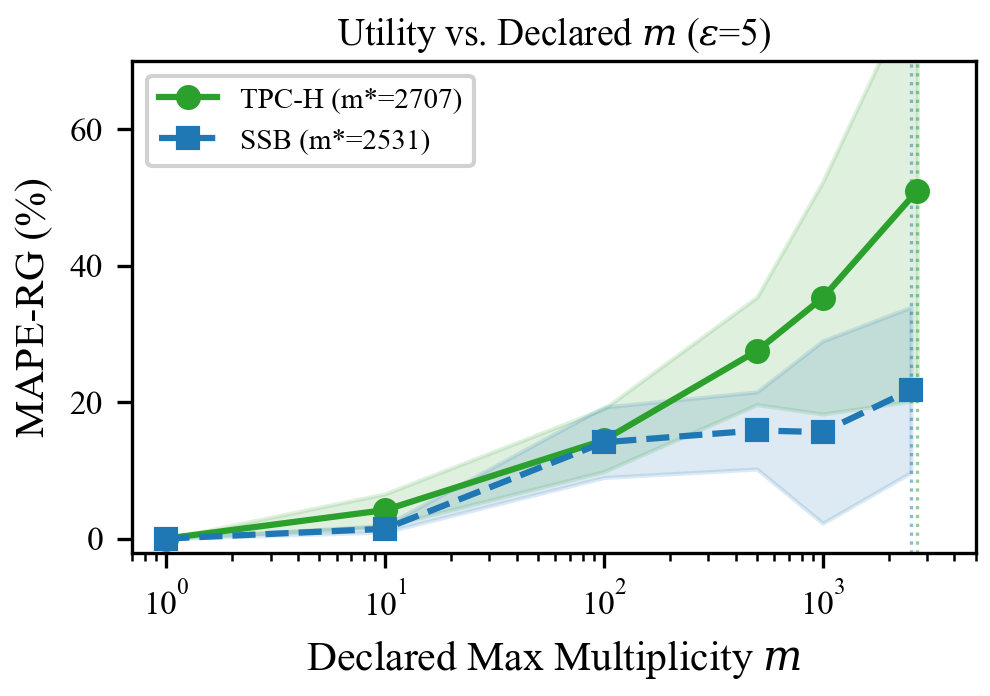}
  \vspace{-1em}
  \caption{MAPE-RG vs.\ declared $m$ at $\varepsilon{=}5$.}
  \label{fig:m_sensitivity}
\end{figure}

Figures~\ref{fig:epsilon} and~\ref{fig:m_sensitivity} use \sys-Full; shaded bands show mean $\pm$ std over five seeds. Figure~\ref{fig:epsilon} fixes $m_{actual} = 2{,}707$ (TPC-H) and $m_{actual} = 2{,}531$ (SSB) and varies $\varepsilon$; Figure~\ref{fig:m_sensitivity} fixes $\varepsilon{=}5$ and varies $m$.

\smallskip
\noindent\textbf{Privacy budget impact.}
A single aggregate MAPE-RG can be misleading: at low selectivity, an off-by-one error (1$\to$2 RGs) yields 100\% APE. At $\varepsilon{=}5$, MAPE-RG for high-selectivity queries ($>$30\%) is only $6.4\%$ (TPC-H) and $8.3\%$ (SSB), well within practical tolerance. Mid-selectivity (5--30\%) reaches $42.9\%$ and $19.6\%$, while low-selectivity ($<$5\%) is highest at $70.9\%$ and $27.8\%$. As $\varepsilon$ increases, all bands converge to 0\% at $\varepsilon{=}\infty$. SSB outperforms TPC-H despite similar domain widths ($\approx$2{,}553 days) and noise scales. The gap reflects TPC-H's more variable per-RG boundary spacing (coefficient of variation $0.25$ vs.\ $0.13$), making tightly packed boundaries more susceptible to noise displacement at low and mid selectivities. Workload-proportional budget allocation is a natural extension that we leave to future work.

\smallskip
\noindent\textbf{Sensitivity to $m$ (Figure~\ref{fig:m_sensitivity}).}
When $\varepsilon{=}5$, MAPE-RG grows as $m$ increases. At $m{=}10$, the error is $4.2\%$ on TPC-H and $1.5\%$ on SSB. At $m{=}100$, the errors rise to $14.5\%$ and $14.1\%$. This happens because the noise scale grows with sensitivity $\Delta_\beta = \min(m,\, d_U - d_L)$, so a higher multiplicity bound increases boundary perturbation until the cap binds at the domain width. If domain knowledge allows for a small $m$, \sys still provides strong utility even when $\varepsilon$ is low.

\subsection{Engine Validation}
\label{sec:eval-duckdb}

Table~\ref{tab:engine} shows median scan times per selectivity band at $\varepsilon{=}5$. High-selectivity queries are well-reproduced by \sys, consistent with their low MAPE-RG of 6.4\% and 8.3\%. Low-selectivity queries exhibit higher MAPE-RG values but only minor absolute timing differences (e.g., 1.4\% on TPC-H), since scanning one or two RGs consistently takes only a few milliseconds. Consequently, a $\pm$1 RG error has minimal impact on wall-clock time at this scale. These results confirm that \sys reproduces scan timing for high-selectivity queries, the regime where I/O cost is largest and performance regressions have the greatest impact.

\begin{table}[t]
  \centering
  \small
  \caption{DuckDB scan times at $\varepsilon{=}5$. (ms)}
  \vspace{-0.5em}
  \label{tab:engine}
      \begin{tabular}{@{}l rrr rrr@{}}
    \toprule
    & \multicolumn{3}{c}{\textbf{TPC-H}} & \multicolumn{3}{c}{\textbf{SSB}} \\
    \cmidrule(lr){2-4}\cmidrule(lr){5-7}
    \textbf{Band} & \textbf{Orig.} & \textbf{\sys} & \textbf{Diff.}
                  & \textbf{Orig.} & \textbf{\sys} & \textbf{Diff.} \\
    \midrule
    Low ($<$5\%)   & 3.59  & 3.54  & $1.4\%$  & 4.63  & 4.22  & $8.8\%$  \\
    Mid (5--30\%)  & 5.34  & 5.12  & $4.2\%$  & 5.90  & 5.70  & $3.3\%$  \\
    High ($>$30\%) & 13.37 & 13.44 & $0.5\%$  & 12.33 & 12.17 & $1.4\%$  \\
    \bottomrule
    \end{tabular}
  \end{table}

\section{Related Work}
\label{sec:related}

\noindent\textbf{Query performance regression diagnosis.}
APOLLO~\cite{jung2020apollo} automates regression detection in DBMSs through domain-specific fuzzing and statistical debugging.
Other recent work~\cite{wu2025understanding} has used pattern-based detection to see how structural changes in query plans affect performance after index tuning. These methods assume access to engine internals or execution traces for diagnosis. \sys addresses a complementary scenario. The developer cannot access the original data at all. Instead, we show how to reproduce the physical I/O behavior using only a privacy-protected sketch.

\noindent\textbf{Privacy-preserving benchmarks.}
PrivBench~\cite{ge2024privacyenhanced} uses sum-product networks to ensure the synthetic data matches the original distribution and query runtime. More broadly, methods based on marginals~\cite{mckenna2022aim}, graphical models~\cite{zhang2017privbayes}, and GANs~\cite{xie2018dpgan} target statistical utility or downstream ML accuracy. These approaches treat the data as a logical object and write it to storage in whatever layout the system chooses. \sys differs fundamentally: we treat the physical storage layout as the target, reproducing per-RG zone maps and compressed sizes rather than value distributions.

\section{Conclusion}
\label{sec:conclusion}

We presented \sys, a framework that reproduces scan-level engine behavior using a privacy-protected Parquet footer sketch. Our results on TPC-H and SSB demonstrate that capturing physical layout metadata, rather than just data distributions, is key to matching engine behavior, achieving 0\% pruning error at $\varepsilon{=}\infty$. We provide pure $\varepsilon$-DP guarantees via the bounded Laplace mechanism, with sensitivity capped at the public domain width. Under DP noise, utility is governed by the multiplicity bound $ m$. For datasets with low multiplicity, \sys achieves strong pruning fidelity even at small $\varepsilon$, while high-selectivity queries remain well-reproduced across both benchmarks at $\varepsilon{=}5$. Tighter sensitivity bounds via DP histograms are a promising direction for future work, alongside extending support to multi-column predicates and downstream operators like joins.

\begin{acks}
This work was supported in part by NSF awards CNS-1846447 and CNS-2016240.
\end{acks}

%\clearpage

\bibliographystyle{ACM-Reference-Format}
\bibliography{references}

\end{document}